\documentclass[english]{article}
\usepackage[T1]{fontenc}
\usepackage[latin9]{inputenc}
\usepackage{float}
\usepackage{graphicx}
\usepackage{babel}
\begin{document}
\title{Mechanism of Universal Quantum Computation in the Brain}
\author{Aman Chawla and Salvatore Domenic Morgera}
\maketitle
\begin{abstract}
In this paper the authors extend \cite{chawla2022gedanken} and provide
more details of how the brain may act like a quantum computer. In
particular, positing the difference between voltages on two axons
as the environment for ions undergoing spatial superposition, we argue
that evolution in the presence of metric perturbations will differ
from that in the absence of these waves. This differential state evolution
will then encode the information being processed by the tract due
to the interaction of the quantum state of the ions at the nodes with
the `controlling' potential. Upon decoherence, which is equal to a
measurement, the final spatial state of the ions is decided and it
also gets reset by the next impulse initiation time. Under synchronization,
several tracts undergo such processes in synchrony and therefore the
picture of a quantum computing circuit is complete. Under this model,
based on the number of axons in the corpus callosum alone \cite{evangelouetal2000}, we estimate
that upwards of 15  million quantum states might be prepared and evolved
every millisecond in this white matter tract, far greater processing than
any present quantum computer can accomplish.
\end{abstract}

\section{Introduction: a Gedanken Experiment}

As per Tegmark \cite{tegmark2000importance}, the smallest neuron
imaginable, with only a single ion traversing the cell, would have
a decoherence time of $10e-14$ seconds. As per \cite{chawla2022computer},
gravitational waves with strains of the order of $h=1e-16$ will have
an impact on axon tract information processing at the time order of
$10e-14$ seconds to $10e-18$ seconds. Multiply each of the above
three time periods by $1e17$ to get 1000 seconds, 1000 seconds and
1 second. Thus, if I start my clock now and a neuron can stay coherent
for 1000 seconds, then in parallel I can consider the other two numbers.
That is, I will send along (in the past) a gravitational wave in the
direction of my neuron tract. Say at $t=0$ it interacts with the
axon tract. As a result, if in its absence there was 0 time difference
between action potential initiation times on the two axons, in its
presence this increases to say 100 seconds. As information is time
coded in the brain, the organism can sense the fact that there is
some timing differential in progress. But can it do something with
this information? Well, suppose the wave is switched on and off 3
times, then about 300 seconds of time differential is accumulated
by the tract input-output. But each of the neurons in the tract was
in a quantum state during this time. This implies that the tract itself
was in a joint quantum state. Now in the absence of these 3 switches,
the quantum state would be preserved, but due to the 3 switches, it
would get perturbed to a different state. This transformation can
then be measured quantum mechanically, in order to detect the passage
of 3 switches of a gravitational wave. If no measurement is performed,
then because the coherence time is an additional 700 seconds, the
perturbed quantum state would in principle be held for another 700
seconds. In contrast if the coherence time was say 0.1 second, then
there would be no initial quantum state which could be perturbed 3
times gently to another quantum state and we would strictly be in
the classical domain. 

The perturbed quantum state is held for 700 seconds. This quantum
information can, if it is part of a larger tract quantum state, be
considered as a new state of the larger tract. In other words, the
gravity wave induced a ``quantum operation'' on part of the brain.
Several such different gravity waves can impinge sequentially or in
parallel in different brain regions, carrying out a quantum computation
in the brain. The moment the computation is over, say, the coherence
time elapses and a decohering mechanism such as a measurement is carried
out on the brain. This measurement thus reads out the result of this
quantum computation. In other words, the brain can act like a quantum
computer \cite{chawla2022gedanken}.

Our paper is structured as follows. In Section \ref{sec:Tegmark-Review}
we provide a review of Tegmark's paper and its implications and differences
as compared to the present work. In Section \ref{sec:Mechanism} we
look in detail at the action potential level, going further down to
the quantum level. Finally, we conclude in Section \ref{sec:Limitations-and-Future}.

\section{Review of Prof. Max Tegmark's Results\label{sec:Tegmark-Review}}

Tegmark's paper contains decoherence computations. He divides the
problem into ion-ion, ion-water and other decoherence mechanisms.
For each such mechanism, he considers a superposition of ions between
the inside and the outside of the cell. Within say the ion-ion collision
mediated decoherence section, he computes how long the spatial superposition
state will stay quantum mechanical. His estimates yield the conclusion
that decoherence precludes a significant role for quantum features
because the characteristic time of a neuron is the action potential
duration of 2 ms, according to him. He also considers microtubules
and the associated decoherence times. 

This absence of quantum effects can be used to argue that the brain
and conscious processes can be entirely explained in terms of classical
physics. In the present paper however, we show that the entire universe
in a sense has a say in the quantum computation being performed in
each part of the brain and the action potential is a very `summed
up' or `gross' indication of what is happening at the deeper physical
layers of the brain\footnote{As opposed to the data layer \cite{chawlamorgerasnider2020}.}.

\section{Mechanism of Quantum Computation\label{sec:Mechanism}}

\begin{figure}[H]
\includegraphics[width=5in]{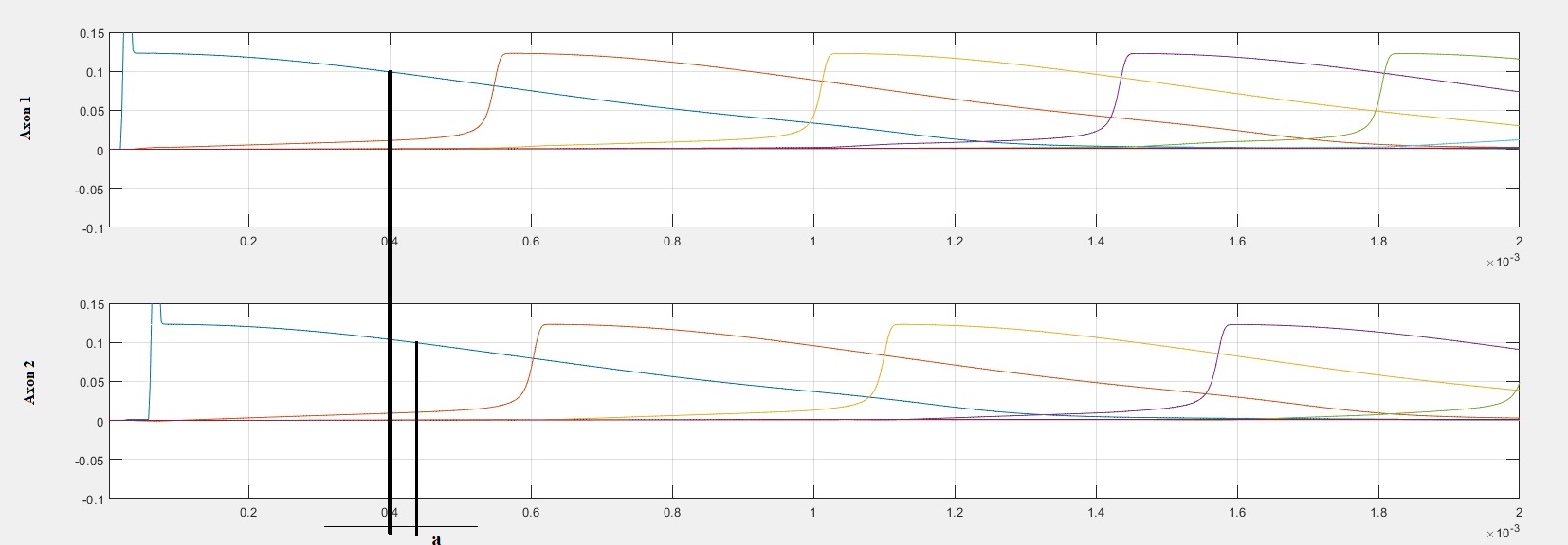}

\caption{Two slightly displaced (in time) action potentials, $V_{1}$ representing
axon 1 (top) and $V_{2}$ representing axon 2 (bottom). The temporal
gap $a$ is indicated at an arbitrary temporal location. \label{fig:Two-slightly-displaced}}

\end{figure}

Consider two slightly temporally displaced action potentials, as shown
in Figure \ref{fig:Two-slightly-displaced}. 

\begin{figure}[H]

\includegraphics[width=5in]{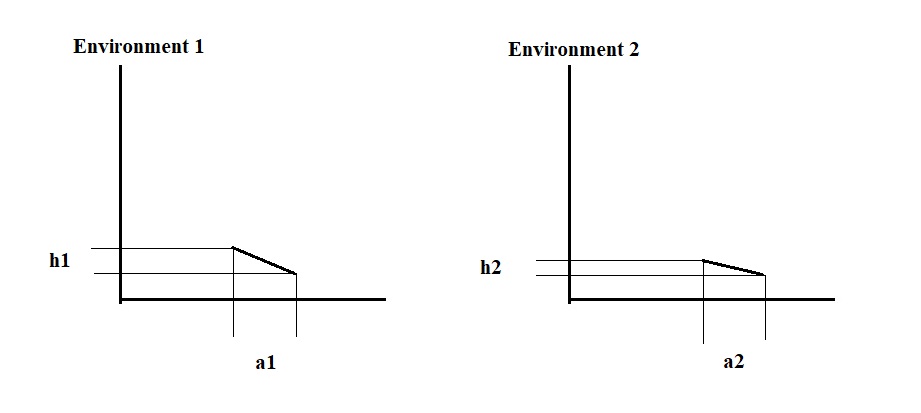}

\caption{Environment in the presence (left) and absence (right) of the gravitational
wave.\label{fig:Environment-in-the}}

\end{figure}

Due to the gravitational wave, the temporal gap between the two action
potentials, $a$, takes the value $a_{1}$ seconds during its presence
and $a_{2}$ seconds during its absence. Suppose that $V_{2}-$$V_{1}$
induces the quantum environment of an ion in the node of Ranvier (please
see Figure \ref{fig:Environment-in-the}). 

\begin{figure}[H]

\includegraphics[width=5in]{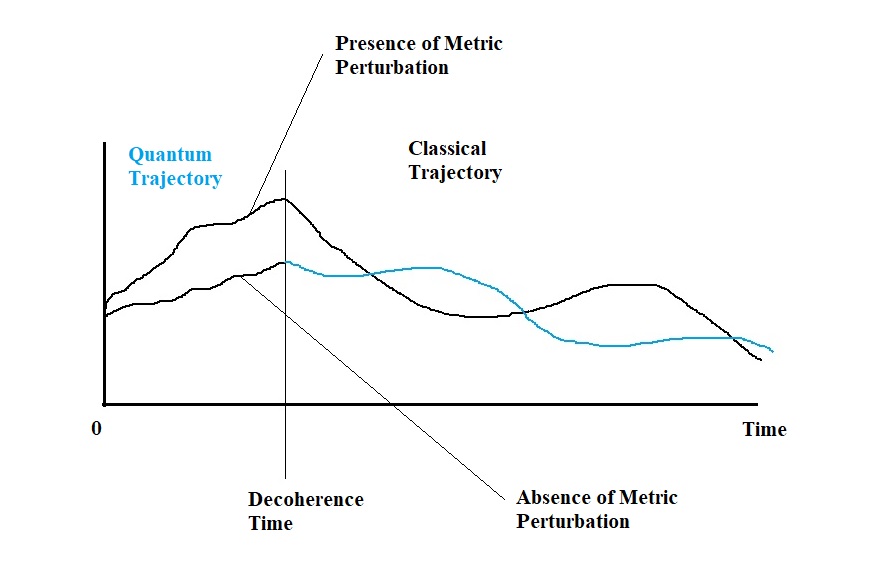}

\caption{Quantum trajectory concatenated with a longer classical trajectory
of the state of the tract with and without the presence of gravitational
radiation. \label{fig:Quantum-trajectory-concatenated}}

\end{figure}

Thus the potential $V(x)$ that enters into the Schrodinger evolution
equation will be different during the presence and the absence of
the gravitational wave. That is, the evolution will be different in
the two cases. Again, if a different gravitational wave-tract interaction
takes place, the evolution will be yet of another type. Post evolution
and subsequent decoherence, and following further classical evolution
(see Figure \ref{fig:Quantum-trajectory-concatenated}), the action
potential initiation process of the cell starts up again and this
re-prepares a quantum state of the tract. 

\begin{figure}[H]
\includegraphics[width=5in]{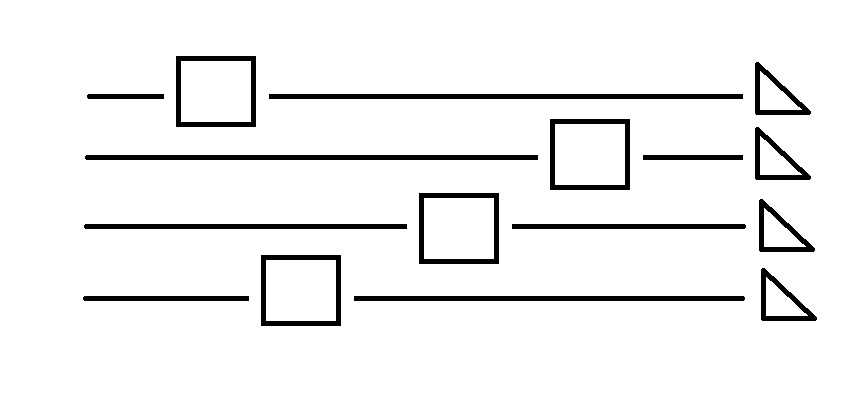}

\caption{Quantum computation in synchronized axon tracts. The square blocks
are where there is interaction with gravitational radiation, resulting
in quantum operations. The triangular blocks at the end represent
the points where the quantum computation comes to an end due to ensuing
decoherence. \label{fig:Quantum-computation-in}}
\end{figure}

Suppose there are a few tracts, each with synchronized impulses, where
the synchronization has taken place using a combination of electric
fields (between tracts) and currents (intratract) \cite{chawla2017axon}.
At a particular instant say $t_{a}$, all the axons are synchronized,
and a joint quantum state is prepared on each of the axons. If there
is a way for us to show that this joint state is actually entangled,
then it results in more interesting processes taking place in the
tract. But regardless, the joint quantum state of the ions will evolve
under the influence of slightly different gravitational impact on
each axon. And because there is this parallel evolution, we have the
process of a quantum circuit evolution taking place. This is illustrated
in Figure \ref{fig:Quantum-computation-in}. 

\section{Limitations and Future Work\label{sec:Limitations-and-Future}}

In this paper, we could have set up a mathematical framework for the
entire process of quantum computation, but did not do so. We felt
that there are adequate presentations of quantum computation in the
literature. Our main goal was to present the novel aspect, namely
that these computations can take place in the coherence window and
the external `control' or `direction' is provided by impinging metric
perturbations. In future work we need to investigate and demonstrate
a mechanism for entanglement between the various tracts.

To conclude, this paper has pushed forward our understanding of how
man and his universe are integrated as one inseparable whole, even
at the level that at least some aspects of what man thinks and does
are enabled and directed by fundamental phenomena in the universe.

\bibliographystyle{unsrt}
\bibliography{mylib2}

\end{document}